\documentclass[%
 reprint,
superscriptaddress,
showpacs,
 amsmath,amssymb,
 aps,prl,
]{revtex4-1}

\usepackage{graphicx}
\usepackage{dcolumn}
\usepackage{bm}
\usepackage{color}
\usepackage{hyperref}

\begin{document}

\title{Experimental Demonstration of Effective Medium Approximation Breakdown\\
in Deeply Subwavelength All-Dielectric Multilayers}

\author{Sergei V. Zhukovsky}
\email{sezh@fotonik.dtu.dk}
\affiliation{DTU Fotonik -- Department of Photonics Engineering, Technical University
of Denmark, {\O}rsteds Plads 343, DK-2800 Kgs.~Lyngby, Denmark}
\affiliation{ITMO University, Kronverksky pr. 49, St. Petersburg, 197101, Russia}

\author{Andrei Andryieuski}
\affiliation{DTU Fotonik -- Department of Photonics Engineering, Technical University
of Denmark, {\O}rsteds Plads 343, DK-2800 Kgs.~Lyngby, Denmark}

\author{Osamu Takayama}
\affiliation{DTU Fotonik -- Department of Photonics Engineering, Technical University
of Denmark, {\O}rsteds Plads 343, DK-2800 Kgs.~Lyngby, Denmark}

\author{Evgeniy Shkondin}
\affiliation{DTU Fotonik -- Department of Photonics Engineering, Technical University
of Denmark, {\O}rsteds Plads 343, DK-2800 Kgs.~Lyngby, Denmark}
\affiliation{DTU Danchip -- National Center for Micro- and Nanofabrication, Technical University
of Denmark, {\O}rstads Plads 347, DK-2800 Kgs.~Lyngby, Denmark}

\author{Radu Malureanu}
\affiliation{DTU Fotonik -- Department of Photonics Engineering, Technical University
of Denmark, {\O}rsteds Plads 343, DK-2800 Kgs.~Lyngby, Denmark}

\author{Flemming Jensen}
\affiliation{DTU Danchip -- National Center for Micro- and Nanofabrication, Technical University
of Denmark, {\O}rstads Plads 347, DK-2800 Kgs.~Lyngby, Denmark}

\author{Andrei V. Lavrinenko}
\email{alav@fotonik.dtu.dk}
\affiliation{DTU Fotonik -- Department of Photonics Engineering, Technical University
of Denmark, {\O}rsteds Plads 343, DK-2800 Kgs.~Lyngby, Denmark}


\begin{abstract}
We experimentally demonstrate the effect of anomalous breakdown of the effective medium approximation in all-dielectric deeply subwavelength thickness ($d \sim\lambda/160-\lambda/30$) multilayers, as recently predicted theoretically [H.H.~Sheinfux et al., Phys.~Rev.~Lett.~113, 243901 (2014)]. Multilayer stacks are composed of alternating alumina and titania layers fabricated using atomic layer deposition. For light incident on such multilayers  at angles near the total internal reflection we observe pronounced differences in the reflectance spectra for structures with 10-nm versus 20-nm thick layers, as well as for structures with different layers ordering, contrary to the predictions of the effective medium approximation. The reflectance difference can reach values up to 0.5, owing to the chosen geometrical configuration with an additional resonator layer employed for the enhancement of the effect. Our results are important for the development of new high-precision multilayer ellipsometry methods and schemes, as well as in a broad range of sensing applications.
\end{abstract}

\pacs{78.67.Pt, 78.20.-e, 42.25.Gy, 81.15.Gh}

\maketitle

\newcommand{\alu}{$\textrm{Al}_2\textrm{O}_3$ }
\newcommand{\tia}{$\textrm{Ti}\textrm{O}_2$ }
\newcommand{\SiN}{$\textrm{Si}_3\textrm{N}_4$ }


Photonic multilayers are one of the most widely studied systems in the broader topic of optics of inhomogeneous media \cite{Brekhovskikh1980,Born1999,yeh1988optical,macleod2010thin}. Most optical effects of such multilayers arise from interference effects underlying the photonic band gap phenomena \cite{Joannopoulos2008}, and therefore are traditionally associated with multilayers, where layer thickness $d$ is comparable to the wavelength of light $\lambda$. For example, the well-known Bragg mirror comprising a stack of alternating low- and high-index dielectric layers, exhibits maximum reflectance if the optical thickness of each layer is close to $\lambda/4$ \cite{Chigrin1999a}. 

From this point of view, the case of multilayers with much thinner layers with thicknesses $d\ll\lambda$ was traditionally regarded as nearly trivial. Indeed, in such a ``deeply subwavelength'' structure the field variation inside a single layer should be very small, leading to negligibly weak interference effects. Therefore, one used to  assume that a light wave interacts with the structure as a whole rather than with its individual layers. The structure can thus be treated as a piece of homogeneous uniaxial material characterized by effective parameters \cite{Simovski2011a}. The applicability of this homogenization approach to all-dielectric multilayers with ultrathin layers $d\ll\lambda$ has always been undoubted, in much the same way as ordinary materials are treated as homogeneous media despite having atomic, molecular, or any other intrinsic structure. 


%
\begin{figure}[b!]
\includegraphics[width=0.95\columnwidth]{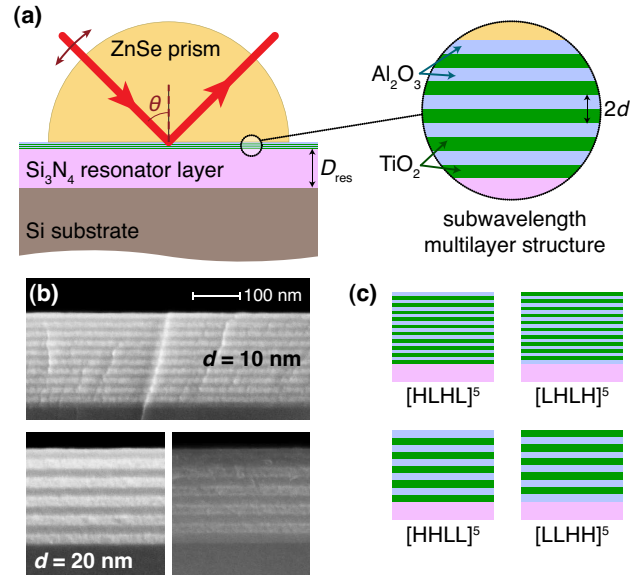}
\caption{(Color online.) (a) Layout of the experimental set-up;
%
%
(b) SEM images of the cross-section of ALD-fabricated multilayers with layer thicknesses 10 nm and 20 nm (with different layer ordering). (c) Schematics of the four fabricated samples differing by the layer thicknesses and ordering.\label{FIG:Structure}}
\end{figure}

However, a recent paper by Sheinfux \textit{et al}. \cite{Sheinfux} showed theoretically that this commonly believed assumption may fail in certain circumstances. Namely, when light is incident on a multilayer at an angle close to that of the total internal reflection (TIR), the actual multilayer and its effective-medium approximation (EMA) model can have significantly different transmission spectra despite the layer thicknesses smaller than $\lambda/50$. Moreover, it was shown that the spectra become sensitive to variations of $d$ on the scale of one nanometer (i.e., $<\lambda/500$), as well as on the layer ordering (i.e., the multilayer HLHL...HL versus LHLH...LH, where H and L stand for high- and low-index layers). Both effects are totally contrary to the predictions of the EMA, which is independent of individual layer features. This ‘anomalous’ EMA breakdown is both enlightening and practically promising (e.g., for sensing and switching applications), therefore it would be of great importance to observe this effect experimentally. 
However, the effect reported in \cite{Sheinfux} only becomes noticeable in multilayers containing at least a few hundreds of 10-nm thick layers with $<1$ nm thickness tolerance, which is extremely challenging to fabricate.

In this Letter, we report the experimental observation of the EMA breakdown effect by placing subwavelength multilayers on a photonic resonator set-up specially designed to enhance reflectance differences arising from the EMA breakdown \cite{ourEMAnano15}. The resonator makes the effect measurable for structures containing only a few tens of layers significantly relaxing the fabrication requirements. The multilayers themselves were fabricated using atomic layer deposition (ALD), chosen for its several advantages compared to other deposition techniques (e.g.~precise thickness control, excellent step coverage, and conformal deposition \cite{ALD}). Our measurements confirm that the reflection spectra of the subwavelength multilayers become sensitive to both layer thickneses and layer ordering in a range of incident angles immediately preceding the TIR angle, in full agreement with the theoretical predictions. 

The layout of the experimental arrangement is shown in Fig.~\ref{FIG:Structure}(a). The central element of the set-up is a subwavelength multilayer film with total thickness 200 nm, containing alternating low-index (L) layers made of alumina ($\textrm{Al}_2\textrm{O}_3$) and high-index (H) layers made of titania ($\textrm{Ti}\textrm{O}_2$). The layer of silicon nitride ($\textrm{Si}_3\textrm{N}_4$) beneath the multilayer serves as an index matched layer 
($n_\text{SiN}\simeq \sqrt{(n_H^2+n_L^2)/2}$) and a resonator required to observe the EMA breakdown. Since the incident angles of interest have to be near-TIR, a high-index ambient medium is required. For this purpose a  semi-cylindrical prism made of zinc selenide ($n_\text{ZnSe}>n_H>n_L$) was used.


All the samples were prepared and assembled in a class 100 cleanroom. The fabrication of the multilayers was performed in a hot-wall ALD system (Picosun R200). The precursors used for \alu  and \tia  deposition were trimethylaluminum $\textrm{Al}(\textrm{CH}_3)_3$ and titanium tetrachloride $\textrm{Ti}\textrm{Cl}_4$, respectively (both from Sigma-Aldrich). In both processes the oxidant source was deionized water. The deposition temperature was 120$^\circ$C in order to prevent the crystal anatase phase transition of \tia  known to occur at temperatures above 150$^\circ$C \cite{TIA2010} which increases the films roughness. 
The growth rates of \alu  and \tia  films were determined to be 0.047 and 0.089 nm/cycle respectively (in agreement with previously reported data \cite{Rate2005}) using varying-time deposition with ellipsometric characterization of the films thicknesses and refractive indices (VASE, J.A. Woollam Co.).

The index matched resonator layer was fabricated by depositing 1040 nm of \SiN  on 100 mm Si(100) wafers using low-pressure chemical vapor deposition. The process was carried out at 780$^\circ$C with ammonia ($\textrm{NH}_3$) and dichlorsilane ($\textrm{SiH}_2\textrm{Cl}_2$) as reactive gases. The thickness and refractive index of the  deposited \SiN film were tested using spectroscopic ellipsometry. 
The film was carefully inspected for cracks, particles and other defects using dark field optical microscopy. The wafer with the best-quality  \SiN coating was selected and cleaved in pieces, which were used as substrates for the subsequent deposition of \alu/\tia multilayers. Before inserting each substrate into the ALD reactor, it was placed on a Si carrier wafer. Therefore the \alu/\tia multilayers were grown not only on the \SiN layer but also on the dummy carrier wafer. After the ALD process, the dummy wafer was cleaved, and its cross-section was characterized using scanning electron microscopy (SEM). The SEM images reveal high-quality homogeneous, conformal coatings, as seen in Fig.~\ref{FIG:Structure}(b).

We fabricated four different configurations of multilayers arrangement, shown schematically in Fig.~\ref{FIG:Structure}(c). Two of them comprise 20 alternating \tia and \alu layers with $d=10$ nm thickness and different layer ordering, i.e., whether the layer closest to the substrate is a \tia or an \alu layer. 
The other two samples comprise 10 alternating layers with double thickness $d=20$ nm, likewise with different layer ordering.  
Note that in the symbolic representation, where H and L refer to 10 nm thick titania and alumina layers, respectively, these four samples can be unambiguously denoted as
$[\mathrm{HLHL}]^{5}$, $[\mathrm{LHLH}]^{5}$, $[\mathrm{HHLL}]^{5}$, and $[\mathrm{LLHH}]^{5}$ [see Fig.~\ref{FIG:Structure}(c)].

\begin{figure}[b]
\includegraphics[width=1.0\columnwidth]{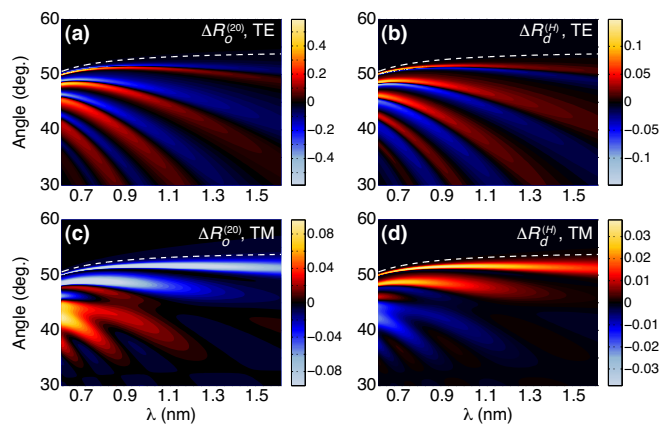}
\caption{(Color online.) Calculated reflectance differences (a) $\Delta R^{(20)}_{o}(\lambda,\theta)$, (b) $\Delta R_{d}^{(H)}(\lambda,\theta)$ for TE-polarized light at near-TIR angles of incidence; (c--d) Same as (a--b) but for TM-polarized light. The dashed line shows the spectral position of the TIR angle $\theta_\text{TIR}(\lambda)$ in presence of material dispersion.
\label{FIG:Theory}}
\end{figure}

If the EMA was valid, all four samples would be homogenized to the same $D=200$ nm thick slab of anisotropic material with permittivity tensor components
\begin{equation}
\epsilon_\parallel=\frac{1}{D}\sum_{j=1}^N{dn_j^2},\quad
\frac{1}{\epsilon_\perp}=\frac{1}{D}\sum_{j=1}^N{dn_j^{-2}},
\label{eq:effmed}
\end{equation}
in the directions parallel and perpendicular to the layers, respectively. Here, $N$ is the number of layers and $n_j$ is the refractive index of the $j^\text{th}$ layer \cite{Rytov1956}. Therefore, the EMA predicts that all samples feature identical reflection spectra. Thus, the difference in the reflectances between the samples signifies the breakdown of the EMA. 

Quantitatively, this breakdown can be analyzed by determining the following two quantities: 
\begin{equation}
\begin{gathered}
\Delta R_d \equiv R|_{d=10\text{ nm}}-R|_{d=20\text{ nm}},\\
\Delta R_o \equiv R|_{\mathrm{H}\ldots}-R_{\mathrm{L}\ldots},
\end{gathered}
\label{eq:DeltaRdRo}
\end{equation}
which reflect the sensitivity of the reflectance spectra towards layer thickness $d$ and layers ordering, respectively. Specifically for the fabricated samples, we can introduce
\begin{equation}
\begin{gathered}
\Delta R_d^{(H)}=R_{[\mathrm{HLHL}]^5}-R_{[\mathrm{HHLL}]^5},\\
\Delta R_d^{(L)}=R_{[\mathrm{LHLH}]^5}-R_{[\mathrm{LLHH}]^5};\\
\Delta R_o^{(20)}=R_{[\mathrm{HHLL}]^5}-R_{[\mathrm{LLHH}]^5},\\
\Delta R_o^{(10)}=R_{[\mathrm{HLHL}]^5}-R_{[\mathrm{LHLH}]^5}.
\end{gathered}
\label{eq:DeltaRspec}
\end{equation}

Figure \ref{FIG:Theory} presents the calculated theoretical dependencies of $\Delta R_d$ and $\Delta R_o$ on the wavelength and angle of plane wave incidence for both polarizations of light (TE and TM). The calculations were performed by the standard transfer matrix approach as used in \cite{ourEMAnano15} with refractive indices of the materials either measured by spectroscopic ellipsometry (\alu, \tia, and \SiN), or obtained from literature (silicon \cite{Si1,Si2}) and  manufacturer-provided data (zinc selenide \cite{ZnSe}). Thus, dispersion in all materials is accurately taken into account. 

In accordance with our earlier theoretical predictions for plane wave incidence \cite{ourEMAnano15}, Fig.~\ref{FIG:Theory} shows a series of characteristic peaks in both $\Delta R_d$ and $\Delta R_o$ with the shape dependent on polarization. These EMA breakdown peaks occur at incidence angles $\theta$ below $\theta_\text{TIR}(\lambda)$
(see the dashed lines in Fig.~\ref{FIG:Theory}). 
The breakdown effect is generally more pronounced in the TE polarization and is more sensitive to the ordering of layers than to the variation of layers thicknesses. 
The values of $\Delta R$ are seen to decrease for larger wavelengths. However, the typical peak values of $|\Delta R|$ for the target range $\lambda$ = 610--1610 nm are between 0.05 and 0.25, and can even reach 0.5, which is favorable for the measurements. Such high values of what would otherwise be a very weak effect are brought about by the resonator layer beneath the subwavelength structure \cite{ourEMAnano15}. 



%
\begin{figure}
\includegraphics[width=0.9\columnwidth]{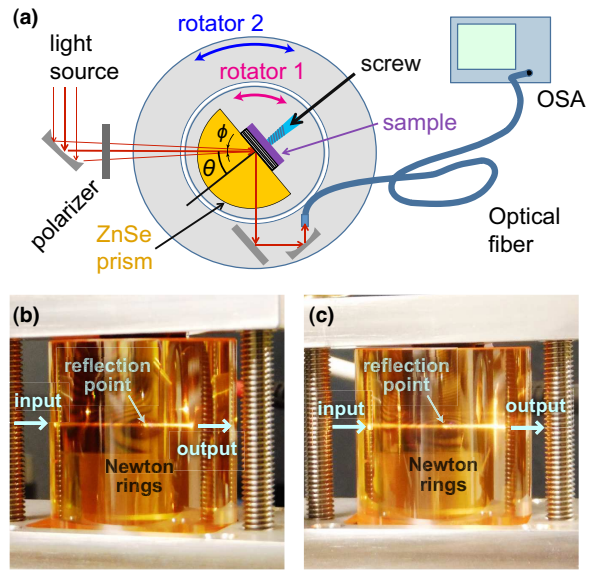}
\caption{(Color online.) (a) Schematics of the experimental set-up for reflectance measurements. 
Photos of (b) 20nm titania terminated sample $[\mathrm{LLHH}]^5$ and (c) 10nm alumina terminated sample $[\mathrm{HLHL}]^5$ affixed to the ZnSe prism taken from the prism side and showing the appearance of the Newton rings, as well as the input and output light beams. \label{FIG:Experimental}}
\end{figure}


To observe the EMA breakdown effect experimentally, we employed a modified Otto-Kretchmann configuration as illustrated in Fig.~\ref{FIG:Structure}(a). 
A multilayer sample is placed in close proximity to a semi-cylindrical ZnSe prism, which is sufficiently high-index to achieve TIR 
at $\theta\sim 50^\circ$.
The light source was a super-continuum broadband  laser (SuperK, NKT Photonics A/S, $\lambda$ = 600--2500 nm). Its collimated output beam was polarized by a double Glan-Thompson polariser and focused at the ZnSe-sample interface, using a set of parabolic mirrors. The reflected beam was collected to a multimode fiber using another parabolic mirror and led into an optical spectrum analyser (OSA, Yokogawa Electric Corp.) with the measuring range $\lambda$ = 350--1750 nm. 

The sample was attached to the prism with a custom-made holder tightened by a small-diameter screw. To minimize the air gap, which would have a dramatic influence on reflectance measurements according to the modelling results (especially for TM polarization), and to reduce the risk of dust trapping between the prism and the sample, the attachment of the sample was performed in the cleanroom. The quality of the attachment was monitored visually controlling the appearance of the Newton rings around the location of the screw on the holder as the screw was tightened. The optical set-up was aligned so that the incident beam was aimed at the innermost region 
of the ring pattern
as shown in Fig. 3(b-c)


Incident angle $\theta$ was varied by a rotating stage with precision 0.17$^\circ$; another rotating stage was used to hold the collecting block (mirrors and fiber) for the reflected beam [Fig.~\ref{FIG:Experimental}(a)]. The focused incident beam was found to span a range $\phi$ of incident angles $\theta$. It was wavelength-dependent and estimated to vary 
from $\phi=1.3^\circ$ at $\lambda=600$ nm to $\phi=3.9^\circ$ at $\lambda=2\,\mu\textrm{m}$.
For each angle $\theta$, intensity spectra were obtained through averaging over
9 independent measurements. The measured spectra were normalized by a reference spectrum obtained by averaging 3 above-TIR spectra ($\theta=53^\circ$, $54^\circ$, $55^\circ$), since $R=1$ for $\theta>\theta_\text{TIR}$. 
The estimated wavelength-averaged error in the reflectance spectra was 4\%.  



%
\begin{figure}
\includegraphics[width=1.0\columnwidth]{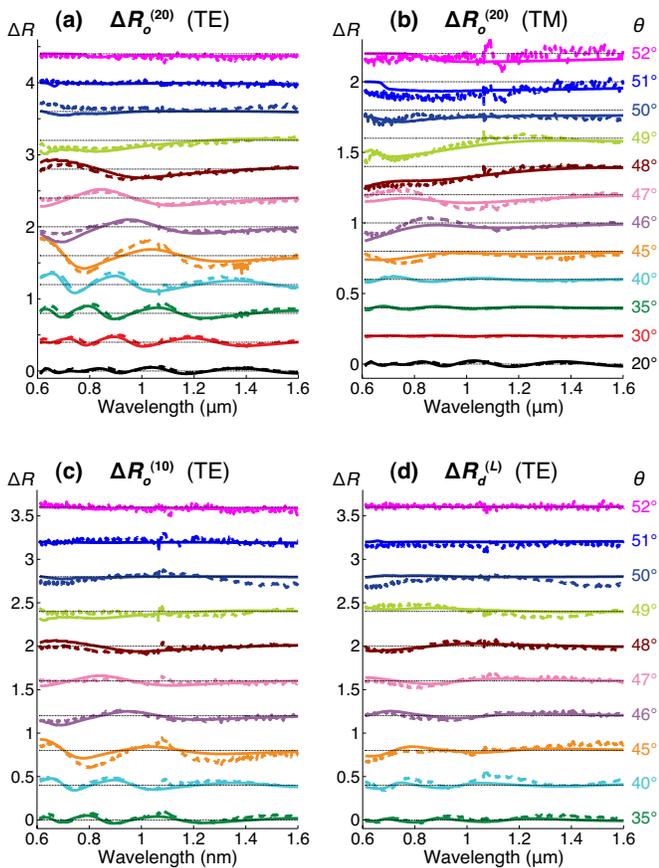}
\caption{(Color online.) Comparison between experimentally measured (dashed lines) and theoretically calculated (solid lines) reflectance differences: $\Delta R_o^{(20)}$ for (a) TE and (b) TM polarization; (c) $\Delta R_o^{(10)}$ and (d) $\Delta R_d^{(L)}$ for the TE polarization. The curves for different incident angles (marked on the right) are shifted vertically by 0.4 (TE) and 0.2 (TM) for the sake of convenience in comparison. 
\label{FIG:Comparison}}
\end{figure}

In order to compare with the experimental measurements, the calculated theoretical spectra were averaged over the wavelength-dependent angle interval $\phi$ corresponding to the beam divergence, as described above. Additionally, a mismatch in $\theta$ equal to $\Delta \theta = 0.2^\circ$ (TE) and $\Delta \theta = 0.3^\circ$ (TM) was introduced to compensate for set-up misalignment. These values of $\Delta \theta$ was determined by minimizing the error between measured and experimental spectra in multivariate optimization (see Supplementary Information). The optimization additionally confirmed that there is no air gap between the prism and the thin-film sample, so an optical contact between them was achieved without the use of immersion liquid (impossible at such high refractive indices).

Both experimental and theoretical results for  $\Delta R_o$ and $\Delta R_d$ in a variety of cases are shown in Fig.~\ref{FIG:Comparison} for a range of $\theta$.  We see that $\Delta R_o^{(20)}$ in the TE polarization [Fig.~\ref{FIG:Comparison}(a)] exhibits small-amplitude ripples across the spectrum for lower values of $\theta$ (20--30${}^\circ$), in line with the expectation of Fig.~\ref{FIG:Theory}(a). As $\theta$ approaches $\theta_\text{TIR}$, the EMA breakdown becomes stronger, and $\Delta R$ reaches values around 0.4--0.5. Above $\theta_\text{TIR}$, $\Delta R$ vanishes since light undergoes TIR ($R=1$) for both samples. Overall, the measured $\Delta R_o^{(20)}$ behaves very close to the theoretical predictions. 

For the corresponding $\Delta R_o^{(20)}$ in the TM polarization 
[see Fig.~\ref{FIG:Comparison}(b)], the reflectance differences have a lower amplitude and fewer characteristic spectral features than for the TE case. However, non-zero $\Delta R$ both in theory and in experiment is still apparent.
%
%
Similar behavior as for $\Delta R_o^{(20)}$, albeit with proportionally smaller amplitudes, is observed for $\Delta R_o^{(10)}$ [Fig.~\ref{FIG:Comparison}(c)], as well as for $\Delta R_d^{(L)}$ [Fig.~\ref{FIG:Comparison}(d)]. There is still a  good agreement between theory and experiment. The measured reflectance difference exceeds the experimental error (4\% average and 3--6\% depending on the wavelength) and reproduces most of the theoretically predicted spectral features.





In summary, we have experimentally demonstrated the effect of the EMA breakdown for all-dielectric multilayers with deeply subwavelength layer thicknesses ($1/30$ to $1/160$ of the incident light wavelength). We have shown that the effect is present even for relatively thin structures, with the total thickness of 200 nm, also smaller than the wavelength. The EMA breakdown manifests as the difference in the reflectance spectra of structures with different layer thickness (20 nm vs.~10 nm) as well as different layer ordering (whether the layer closer to the substrate is a high- or a low-index layer), as shown in Fig.~\ref{FIG:Structure}(c). The measured reflectance difference spectra, reaching values of around 0.5, are in good agreement with theoretical transfer matrix calculations (Fig.~\ref{FIG:Comparison}). 

Our results can be used in ellipsometry of multilayer structures, both to correct the existing ellipsometry models that rely on the EMA, and to devise new models specifically based on the measurement of the features related to the EMA breakdown. 


As regards the applicability of the obtained results to sensing (using the high sensitivity of the EMA breakdown to the incident angle and the refractive index behind the multilayer \cite{ourEMAnano15}), we see that the spectral features in the reflectance spectra are not as sharp as theoretically predicted. This is primarily due to the use of the hemicylindrical prism together with a focused beam from a broadband light source, causing the light waves that reach the sample to have a finite range $\phi$ of incident angles $\theta$ and therefore smearing the spectral features. It is anticipated that adjusting the  experimental set-up to reduce this range of angles (e.g. by using a triangular prism or a light source with a collimated output) would lead to sharper spectral features in the reflectance difference spectra (see Fig.~\ref{FIG:Theory}) suitable for high-sensitivity angle and refractive index measurements.

The authors acknowledge fruitful discussions with A. Novitsky. Partial support from the People Programme (Marie Curie Actions) of the EU 7th Framework Programme FP7-PEOPLE-2011-IIF under REA grant agreement No. 302009 (Project HyPHONE) is gratefully acknowledged. A.A. acknowledges  support from the Danish Council for Independent Research via the GraTer project (Contract No. 0602-02135B).

\bibliographystyle{apsrev4-1}
\bibliography{EMA_breakdown}

\end{document}